\documentclass[sigconf,nonacm]{acmart}

\setcopyright{none}
\title{Beyond Supra-Competitive Outcomes: Collusive Behaviour in Deep Reinforcement Learning for Optimal Execution Games}

\author{Christos S. Koulouris}
\affiliation[obeypunctuation=true]{
  \department{Institute of Finance and Technology}\\
  \institution{University College London}\\
  Gower Street, \city{London} WC1E 6BT, \country{United Kingdom}
}

\author{Carlo Campajola}
\affiliation[obeypunctuation=true]{
  \department{Institute of Finance and Technology}\\
  \institution{University College London}\\
  Gower Street, \city{London} WC1E 6BT, \country{United Kingdom}
}
\affiliation[obeypunctuation=true]{
  \institution{UZH Blockchain Center}\\
  Andreasstrasse 15, 8050 \city{Z\"urich}, \country{Switzerland}
}

\renewcommand{\shortauthors}{Koulouris and Campajola}

\begin{document}

\begin{abstract}
In this paper, we extend earlier findings of supra-competitive outcomes in optimal-execution games by identifying a learned punitive mechanism that deters deviations and provides behavioural evidence of collusion. We investigate this mechanism in a two-player, finite-horizon Almgren--Chriss liquidation game. Independent proximal policy optimisation agents with access to within-episode price and action histories achieve costs below the Nash benchmark. We identify a profitable deviation by training against the mean learned liquidation schedule, then impose its first trade on one of the original agents. The opponent responds by accelerating liquidation. This response more than offsets the deviator's gain in every run and both player roles, while leaving the punisher's average payoff materially unchanged relative to not punishing under the same deviation. The punisher imposes greater losses on the deviator while preserving its own average payoff, despite the availability of more profitable, less punitive liquidation plans. Matching deviations and subsequent additional selling rise and later decline during training, while final policies retain an effective punitive response. We formalise two checks: whether punishment outweighs the gain from deviating, and whether the change in trading behaviour is large enough to account for the loss imposed. Both checks hold for the tested deviation. Together, these findings provide behavioural and economic evidence supporting a collusive interpretation of the learned supra-competitive outcomes.
\end{abstract}


\maketitle

\section{Introduction}
\label{sec:introduction}

Learning algorithms increasingly make economic decisions in environments where their rewards depend on the actions of other adaptive agents. Although each agent is trained to maximise its own reward, simulation studies have observed that such learning may not lead to competitive outcomes. In pricing games, learning agents have been found to sustain supra-competitive profits without explicit communication or a shared objective~\cite{calvano2020artificial,klein2021autonomous}. These findings raise a central question for multi-agent learning: what behaviours allow self-interested agents to sustain such outcomes?

Punishment provides a mechanism through which cooperation can be sustained. A unilateral departure from a cooperative strategy may offer an immediate gain, while the prospect of an adverse response changes the value of that departure over the remainder of the interaction. Harrington~\cite{harrington2018developing} places reward--punishment strategies at the centre of a behavioural account of collusion: cooperation is rewarded, and departures from it are penalised. Finding such responses in learned policies is therefore significant because it connects supra-competitive outcomes to the strategic behaviour that can sustain them, supporting their interpretation as collusive rather than merely supra-competitive.

Calvano et al.~\cite{calvano2020artificial} provide a foundational empirical demonstration in repeated price competition. Their independent tabular Q-learners develop strategies under which imposed unilateral price cuts trigger temporary retaliation, followed by a gradual return towards cooperation. By examining both the responses and the resulting discounted payoffs, they show how punishment can deter deviations from supra-competitive pricing. Klein~\cite{klein2021autonomous} extends the evidence to sequential price competition, where Q-learning agents can learn supra-competitive pricing supported by punishment strategies. Together, these studies make deviation-response experiments a natural way to investigate the mechanisms underlying algorithmic collusion.

This perspective also informs recent work. In a repeated prisoner's dilemma, Bertrand et al.~\cite{bertrand2025selfplay} analyse one-step-memory Q-learning in self-play with a shared Q-table and establish conditions under which learning moves towards Pavlov, a strategy that supports cooperation through contingent punishment and recovery; they also report an empirical deep-Q-network extension. In electricity markets, Seredy\'nski and Tsaousoglou~\cite{seredynski2026electricity} use imposed deviations and best-response experiments to investigate enforcement, identifying temporary punitive bidding followed by recovery in selected market configurations. Sivachandran and Paleja~\cite{sivachandran2026curb} target retaliation directly: their CURB method combines penalties on history-dependent changes in action distributions with synthetic experiences of unpunished deviations. Their pricing and quantity-competition experiments report reduced collusive outcomes, while forced-deviation tests show attenuated retaliatory responses.

Financial trading gives these interactions a distinctive economic structure. In optimal execution, an agent must complete a prescribed trade within a finite horizon while accounting for its own effect on prices. The Almgren--Chriss framework~\cite{almgren2001optimal} formalises the trade-off between market impact and exposure to price risk. Nevmyvaka et al.~\cite{nevmyvaka2006reinforcement} demonstrate how reinforcement learning can adapt execution decisions to inventory, remaining time and market conditions.

Ning et al.~\cite{ning2021double} use Double Deep Q-learning for liquidation subject to inventory constraints, while Karpe et al.~\cite{karpe2020multiagent} apply it to execution in an interactive limit-order-book simulator. Moallemi and Wang~\cite{moallemi2022reinforcement} study reinforcement learning for the timing of individual child orders, while Schnaubelt~\cite{schnaubelt2022deep} compares value-based and policy-gradient methods for cryptocurrency limit-order placement. Macr\`i and Lillo~\cite{macri2024liquidity} use Double Deep Q-learning to adapt execution to deterministic and stochastic changes in price impact without directly observing the impact coefficients. Cheridito and Weiss~\cite{cheridito2026execution} learn dynamic allocations between market and limit orders through a logistic-normal actor--critic policy. Espana et al.~\cite{espana2025queue} train a DDQN execution agent in a queue-reactive limit-order-book simulator, allowing its trades to affect liquidity and subsequent order flow.

When several agents trade the same asset, their execution costs become coupled through market impact: one agent's liquidation schedule changes both its own trading conditions and those faced by its competitors. Schied and Zhang~\cite{schied2017state} formalise this strategic interaction through a state-constrained differential game of optimal liquidation.

This coupling creates scope for both cooperative execution and punitive responses. Jointly moderating the rate of liquidation can reduce the costs generated by competition for liquidity, whereas a response that accelerates trading can worsen the execution opportunities remaining to another agent. Inventory depletion and the liquidation deadline shape the gains from departing from a joint trading pattern and the consequences of an opponent's reaction. Optimal execution therefore offers a setting in which to study how deep reinforcement-learning agents coordinate their actions and whether they learn behaviours that support that coordination.

Two recent studies provide the closest foundations for this investigation. Lillo and Macr\`i~\cite{lillo2026deviations} study two independent Double Deep Q-learning agents liquidating the same asset under market impact. They report execution costs below their Nash benchmark, with learned schedules often closer to cooperative liquidation. Their discussion identifies controlled deviations and subsequent responses as a promising route to understanding whether reward--punishment mechanisms sustain these outcomes. Koulouris and Campajola~\cite{koulouris2026memory} examine the role of information and memory in execution learning, comparing ex-ante schedule selection, state-dependent policies, stronger conditioning on the current price, and a history-aware architecture processing past prices and the agent's own actions. Access to within-episode histories produces more frequent and persistent supra-competitive outcomes, while stronger current-price conditioning alone does not reproduce this pattern. They discuss punishment-like responses as a channel that history-dependent feedback could support, motivating a direct examination of responses to unilateral deviations.

Building on these studies, this paper examines collusive behaviour between deep reinforcement-learning agents in optimal execution games. We observe punitive responses following unilateral deviations from the learned liquidation pattern. We investigate these responses as a mechanism that can help sustain supra-competitive outcomes, providing a behavioural basis for characterising the learned interaction as collusive and connecting the execution literature to the analysis of algorithmic collusion.

\section{Execution Environment and Learning Agents}
\label{sec:environment-agents}

\subsection{The liquidation game}
\label{sec:execution-environment}

We consider a discrete-time Almgren--Chriss liquidation game. Two agents, indexed by $k\in\{1,2\}$, liquidate the same asset over a horizon $T$, divided into $N$ trading periods of length $\tau=T/N$. Agent $k$ starts with inventory $q_0^{(k)}>0$ and zero cash. At period $t\in\{1,\ldots,N\}$, both agents simultaneously submit signed trade quantities $v_t^{(k)}$: positive values denote sales and negative values denote purchases. Writing $V_t=\sum_{k=1}^{2}v_t^{(k)}$ for aggregate flow, inventories evolve as
\begin{equation}
  q_t^{(k)}=q_{t-1}^{(k)}-v_t^{(k)},
  \qquad q_N^{(k)}=0.
  \label{eq:inventory-dynamics}
\end{equation}

The publicly observed mid-price includes accumulated permanent impact. Starting from $S_0$, the execution price and subsequent mid-price are
\begin{align}
  \widetilde S_t
    &=S_{t-1}-a\frac{V_t}{\tau},
    \label{eq:execution-price}\\
  S_t&=S_{t-1}-\kappa V_t+\sigma\sqrt{\tau}\,\xi_t,
  \qquad \xi_t\overset{\mathrm{iid}}{\sim}\mathcal N(0,1),
  \label{eq:midprice-dynamics}
\end{align}
where $a>\kappa>0$ are the temporary- and permanent-impact coefficients, respectively, and $\sigma$ is exogenous volatility. Both agents trade at the same execution price, with temporary impact determined by their combined contemporaneous orders. The random price shock is realised after execution and is unavailable when either action is selected. Cash satisfies $C_0^{(k)}=0$ and $C_t^{(k)}=C_{t-1}^{(k)}+\widetilde S_tv_t^{(k)}$.

\subsection{Execution cost and reward}
\label{sec:reward}

Each agent minimises its own expected implementation shortfall,
\begin{equation}
  \mathrm{IS}^{(k)}
    =S_0q_0^{(k)}-C_N^{(k)}
    =S_0q_0^{(k)}-\sum_{t=1}^{N}
       \widetilde S_tv_t^{(k)}.
  \label{eq:implementation-shortfall}
\end{equation}
The objective is risk-neutral: there is no variance penalty or shared reward. We report $\mathrm{IS}^{(k)}/N$, a fixed rescaling of total shortfall, for which lower values indicate better execution.

The environment supplies the per-period reward
\begin{equation}
  r_t^{(k)}=-\frac{S_0q_0^{(k)}}{N}
                +\widetilde S_tv_t^{(k)}.
  \label{eq:environment-reward}
\end{equation}

\subsection{Equilibrium benchmarks}
\label{sec:benchmarks}

Nash equilibria for optimal-execution games have been characterised by Schied and Zhang~\cite{schied2017state} and by Cordoni and Lillo~\cite{cordoni2024transient}. For the analysis below, we follow the benchmark comparison of Koulouris and Campajola~\cite{koulouris2026memory}, using Nash equilibrium as the competitive reference and time-weighted average price (TWAP) liquidation as the cooperative reference. In our feedback setting, the Nash reference is the finite-grid closed-loop equilibrium. For equal initial inventories $q_0^{(1)}=q_0^{(2)}=q_0$, joint TWAP prescribes $v_t^{(1)}=v_t^{(2)}=q_0/N$ in every period. Lillo and Macr\`i~\cite{lillo2026deviations} prove that this joint liquidation strategy is Pareto efficient in their symmetric, risk-neutral two-player execution game.

\subsection{Information and history encoding}
\label{sec:history-encoding}

Each player has its own actor--critic network, optimiser and on-policy rollout buffer. Its decision input $o_t^{(k)}$ comprises the current mid-price $S_{t-1}$, its own inventory $q_{t-1}^{(k)}$, the period index, and its within-episode price and executed-action histories. Opponent inventories, actions, rewards and network parameters are not supplied to either network. Past opponent trading can affect the observed price history, but both current actions are chosen before the environment advances. Histories reset at the beginning of every episode; no previous episode is supplied as part of the policy input.

Following the history-encoding construction in Koulouris and Campajola~\cite{koulouris2026memory}, prices and own actions enter a single sequence encoder. Under our period indexing, the valid tokens at decision $t$ are $(S_0,0),(S_1,v_1^{(k)}),\ldots,(S_{t-1},v_{t-1}^{(k)})$.

Each token is projected to 64 dimensions and given a sinusoidal positional encoding. A Transformer with two encoder layers, two attention heads per layer, and feed-forward width 256 processes the sequence. Unobserved positions are masked, and the output is averaged over valid positions and layer-normalised. Attention can connect all observed positions within the current prefix; future prices and trades are excluded.

In parallel, a two-layer, width-64 multilayer perceptron processes the normalised current price, own inventory and period index. A two-layer, width-128 network combines this representation with the pooled history. Separate scalar heads produce the actor's residual latent mean and the critic's value estimate. The networks use GELU activations and layer normalisation, with zero dropout. Actor and critic share this representation within each agent; the two agents share no trainable parameters.

\subsection{Continuous-action PPO}
\label{sec:ppo}

We train independent policies using the clipped-surrogate version of proximal policy optimisation (PPO)~\cite{schulman2017ppo}. This choice reflects the game's inherently continuous action space: PPO directly parameterises a policy over trade quantities, in contrast to the value-based double deep Q-learning approaches adopted by Lillo and Macr\`i~\cite{lillo2026deviations} and Koulouris and Campajola~\cite{koulouris2026memory}. For a non-degenerate feasible action interval, let $m_t^{(k)}$ and $d_t^{(k)}$ denote its midpoint and half-width. The policy samples a scalar latent variable and transforms it into a feasible trade:
\begin{align}
  z_t^{(k)}&\sim\mathcal N\!\left(
      \mu_{\theta_k}(o_t^{(k)}),s_e^2\right),\nonumber\\
  v_t^{(k)}&=m_t^{(k)}+d_t^{(k)}\tanh z_t^{(k)},
  \label{eq:squashed-policy}
\end{align}
where $s_e$ is the exploration scale for the current training rollout. Degenerate intervals, including the final period, produce the uniquely feasible action directly.

Both policies remain fixed while collecting 200 complete episodes, giving 2,000 transitions per agent. Each then updates from its own rollout before further interaction. With $\gamma=1$ and generalised advantage estimation parameter $\lambda_{\mathrm{GAE}}=1$, the unnormalised advantage is
\begin{equation}
  A_t^{(k)}=\widehat R_t^{(k)}-
     \mathcal V_{\theta_k^{\mathrm{old}}}(o_t^{(k)}),
  \label{eq:returns-advantages}
\end{equation}
where $\widehat R_t^{(k)}$ is the Monte Carlo return target and $\mathcal V$ denotes the critic. Advantages are standardised over genuine policy decisions in the rollout. Let $\widehat A_t^{(k)}$ denote these standardised values and let
\[
  \rho_t^{(k)}(\theta_k)=
  \frac{\pi_{\theta_k}(v_t^{(k)}\mid o_t^{(k)})}
       {\pi_{\theta_k^{\mathrm{old}}}(v_t^{(k)}\mid o_t^{(k)})}.
\]
The actor maximises the empirical clipped objective
\begin{equation}
\begin{split}
  L_k^{\mathrm{clip}}(\theta_k)=\widehat{\mathbb E}_{\mathcal D_k}
  \big[\min\{&\rho_t^{(k)}(\theta_k)\widehat A_t^{(k)},\\
    &\operatorname{clip}(\rho_t^{(k)}(\theta_k),1-\epsilon,1+\epsilon)
       \widehat A_t^{(k)}\}\big],
\end{split}
  \label{eq:ppo-clipped-objective}
\end{equation}
where $\mathcal D_k$ contains the non-forced decisions and $\epsilon=0.10$. Ratios are evaluated from the stored Gaussian latents: the state-dependent transformation in~\eqref{eq:squashed-policy} has the same Jacobian under old and new policies, so it cancels. The combined loss is $-L_k^{\mathrm{clip}}+c_V L_k^V$, where $L_k^V$ is one-half the mean squared return-prediction error over all transitions and $c_V=0.5$. Forced final trades are excluded from the actor objective, but their rewards remain in earlier returns and their states remain in critic training. There is no entropy bonus.

Each update uses Adam, shuffled mini-batches and at most four epochs. Early stopping monitors old-to-new policy KL divergence, with threshold $0.005$ over the rollout and $1.5$ times that threshold for the mini-batch estimate; this is a stopping rule rather than a hard trust-region constraint. Rollouts are discarded after each update. The learning rate decreases linearly during training. The latent log-standard-deviation decreases linearly from $-1.2$ to $-4$ over the first $75\%$ of training, then remains fixed; it is held constant throughout each rollout and its optimisation. Exploration draws are independent between agents.

We train ten independent agent pairs for 40,000 episodes each. We evaluate each run over 500 test episodes with frozen parameters.

\section{Supra-Competitive Outcomes}
\label{sec:supra-competitive-results}

We evaluate the ten trained pairs with $q_0^{(1)}=q_0^{(2)}=100$, $N=T=10$, $S_0=10$, $a=0.002$ and $\kappa=0.001$. Throughout the experiments that follow, we set $\sigma=10^{-9}$, following the low-noise setting of Lillo and Macr\`i~\cite{lillo2026deviations}. Exogenous price fluctuations are therefore negligible relative to market impact, making the effects of deviations and subsequent trading responses easier to interpret.

\begin{figure}[tb]
  \centering
  \includegraphics[width=\columnwidth]{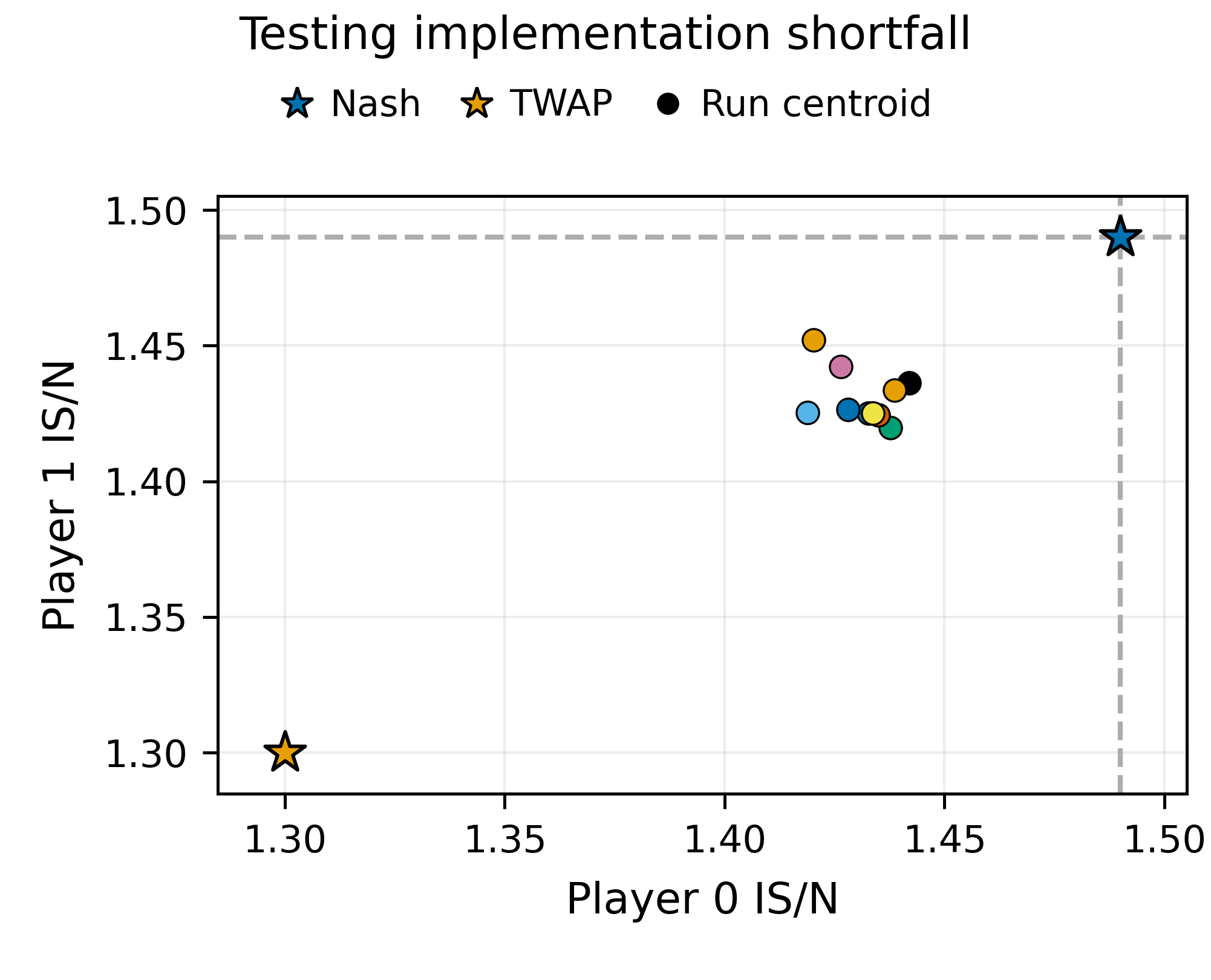}
  \caption{Testing implementation shortfall for the two players. Each circle averages 500 test episodes from one trained pair; stars mark the closed-loop Nash and joint TWAP benchmarks. Lower values on both axes indicate a joint improvement in execution costs.}
  \Description{Ten run centroids lie below the Nash benchmark in both players' implementation shortfall, but above the TWAP benchmark.}
  \label{fig:testing-centroids}
\end{figure}

Figure~\ref{fig:testing-centroids} reports the pair of mean testing costs for each run. Both players achieve lower $\mathrm{IS}/N$ than the closed-loop Nash benchmark in all ten runs. They remain above the joint TWAP benchmark. The learned outcomes are thus supra-competitive, while falling short of the cooperative benchmark. This pattern is qualitatively consistent with the supra-competitive outcomes reported for history-aware agents by Koulouris and Campajola~\cite{koulouris2026memory}. In this setting with aggregate temporary and permanent impact, the 500 test episodes with deterministic action selection within each run produce essentially overlapping cost pairs at the figure's scale. The visible dispersion primarily reflects differences between independently trained pairs.

\begin{figure*}[!t]
  \centering
  \includegraphics[width=\textwidth]{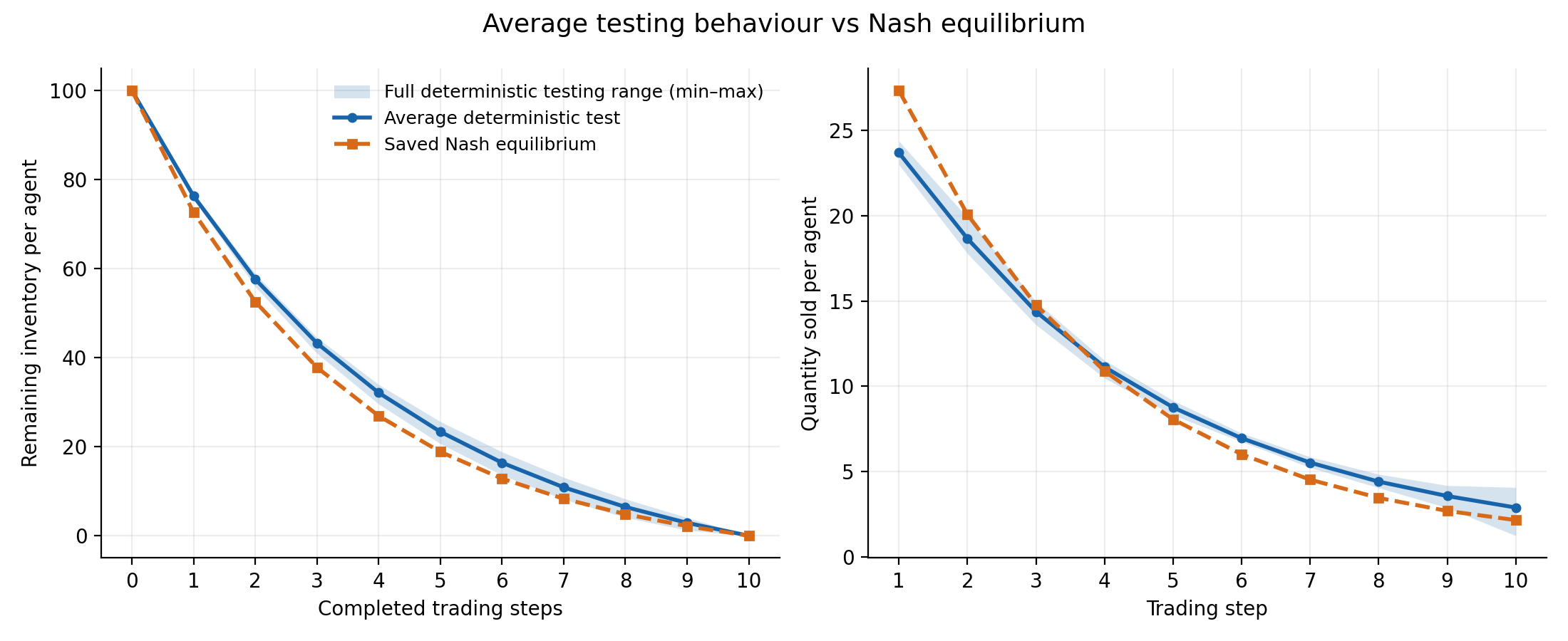}
  \caption{Mean testing inventory paths (left) and trade quantities (right), compared with closed-loop Nash. The shaded region is the full pointwise range across both players, ten trained runs and 500 test episodes per run. It measures realised trajectory dispersion.}
  \Description{Mean inventory and trading paths with a minimum-to-maximum envelope covering every deterministic testing trajectory, compared with the closed-loop Nash schedule. Learned liquidation is slower initially, and the envelope is narrow.}
  \label{fig:average-testing-behaviour}
\end{figure*}

Figure~\ref{fig:average-testing-behaviour} shows the arithmetic mean of the testing inventory paths and trade quantities, pooled across both players and all runs. The shaded region spans the pointwise minimum and maximum over all 10,000 agent trajectories. This shows that the mean represents similar realised liquidation paths, rather than concealing large differences between them. On average, the agents sell less than Nash in the opening periods and retain more inventory throughout the interior of the episode, shifting liquidation towards later periods.

Prior work shows that limited or rapidly declining exploration can itself generate supra-competitive outcomes~\cite{abada2023artificial,abada2024collusion}. Our agents retain stochastic action exploration throughout training, and the Nash liquidation path is feasible under their action constraints. We verify from the training trajectories that each player's actions cover ranges on both sides of the Nash trade quantity at every discretionary step, while joint training costs repeatedly approach the Nash benchmark in all ten runs. These checks demonstrate exploration beyond the final supra-competitive region.

These results establish supra-competitive execution outcomes. The deviation experiments that follow examine whether contingent punitive responses support this behaviour.

\section{Punishing Profitable Deviations}
\label{sec:punishing-profitable-deviations}

\subsection{Learning a Profitable Deviation}
\label{sec:learned-deviation}

The supra-competitive liquidation pattern suggests that a player could reduce its execution cost by deviating while its opponent continues to follow that pattern. We test this conjecture by training a PPO agent against a fixed version of the learned trajectory, with the primary aim of identifying an economically meaningful deviation for the subsequent intervention experiments.

Let $\bar{\mathbf v}=(\bar v_1,\ldots,\bar v_N)$ denote the arithmetic mean of the test trades across both players and all ten self-play runs. The close agreement between the realised paths in Figure~\ref{fig:average-testing-behaviour} supports using this pooled schedule as a representative reference for the observed supra-competitive liquidation. We replace one learning agent with a non-adaptive player that executes $\bar v_t$ at period $t$ in every training and test episode, independently of prices or its opponent's actions. Thus, we fix the opponent's trajectory, rather than merely freezing the parameters of a policy that could still react to deviations.

The remaining player is a newly initialised PPO agent with the same architecture, information, reward and training protocol as in Section~\ref{sec:ppo}. The fixed schedule is not supplied as an additional input to its network. We ensure that the fixed agent fully liquidates its initial inventory by the final trading period. We conduct ten independent training runs of 40,000 episodes and evaluate each learner over 500 test episodes with frozen parameters. The execution environment and low-noise setting remain unchanged.

The relevant comparison holds the opponent's schedule fixed: the learner's cost is compared with the cost of following $\bar{\mathbf v}$ against another copy of $\bar{\mathbf v}$. The learned responses yield a positive reduction in mean testing cost in every run. We then average the learner's test trades across the ten runs to obtain a common deviation trajectory $\mathbf d$; the fixed player's trades are excluded from this average. Evaluation using the environment's expected execution cost confirms that this averaged trajectory also remains profitable against $\bar{\mathbf v}$. Writing $J(\mathbf x,\mathbf y)$ for the expected $\mathrm{IS}/N$ of a player following schedule $\mathbf x$ against schedule $\mathbf y$, we obtain
\begin{equation}
  J(\mathbf d,\bar{\mathbf v})
  < J(\bar{\mathbf v},\bar{\mathbf v}).
  \label{eq:profitable-learned-deviation}
\end{equation}

This profitable unilateral departure establishes that the symmetric fixed-schedule profile $(\bar{\mathbf v},\bar{\mathbf v})$ is not a Nash equilibrium. The following experiments therefore use $\mathbf d$ to construct deviations while restoring the opponent's learned capacity to respond, allowing us to examine whether punitive behaviour counteracts the incentive identified here.

\subsection{Identifying Learned Punitive Behaviours}
\label{sec:learned-punishment}

We return to the ten jointly trained agent pairs and freeze their parameters. In each of the 500 test episodes per pair, we force one player, the \emph{deviator}, to execute $d_1$, the first trade of the deviation trajectory identified in Section~\ref{sec:learned-deviation}. Only this first trade is replaced: from timestep 2 onwards, the deviator resumes its learned policy using its actual inventory and observed history. The other player, which we call the \emph{punisher}, follows its learned policy throughout. We repeat the experiment with the roles reversed. Both agents must fully liquidate by the final timestep.

\begin{figure}[!htbp]
  \centering
  \includegraphics[width=\columnwidth]{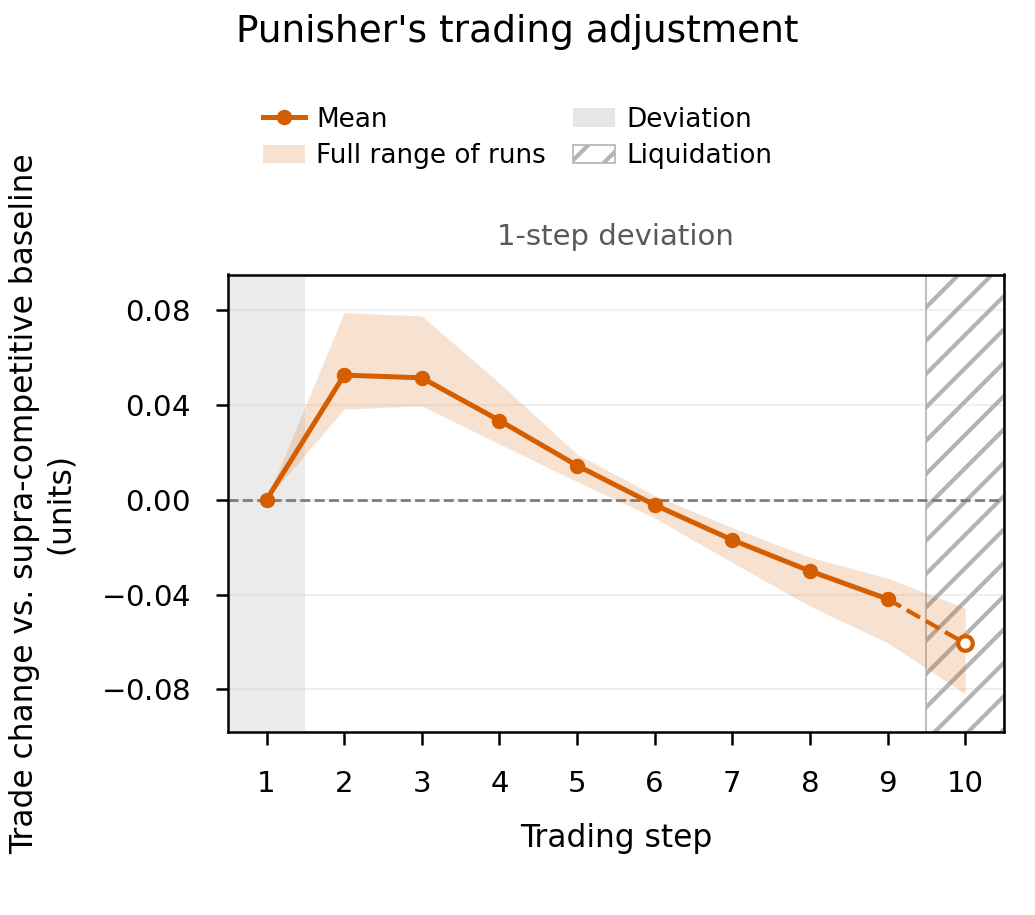}
  \caption{Punisher's trade under the learned response to the deviation minus its trade along the learned supra-competitive path without the deviation. The line averages these differences over both role assignments, 500 paired test episodes and ten trained runs; shading spans the full range of run means. Grey and hatched regions mark the deviation and compulsory final liquidation.}
  \Description{The punisher sells more in timesteps two to five than in the unforced episode, then sells less at later timesteps. Its first trade is unchanged.}
  \label{fig:punishment-action-changes}
\end{figure}

Figure~\ref{fig:punishment-action-changes} shows, at each timestep, the punisher's executed trade under its learned response to the deviation minus its trade along the learned supra-competitive path without the deviation. Because trades are simultaneous, its first action is unchanged; the deviation can affect its decision only from timestep 2, through the observed price history. All ten run means exhibit additional selling in timesteps 2--5, followed by reduced sales later in the episode. The punisher therefore accelerates liquidation after the deviation becomes observable. Through aggregate temporary and permanent impact, earlier selling depresses execution prices while the deviator still holds inventory. This pattern is consistent with a punitive response that exposes the deviator to earlier price impact. The adjustment at the final timestep reflects compulsory liquidation of the remaining inventory.

\begin{figure}[tb]
  \centering
  \includegraphics[width=\columnwidth]{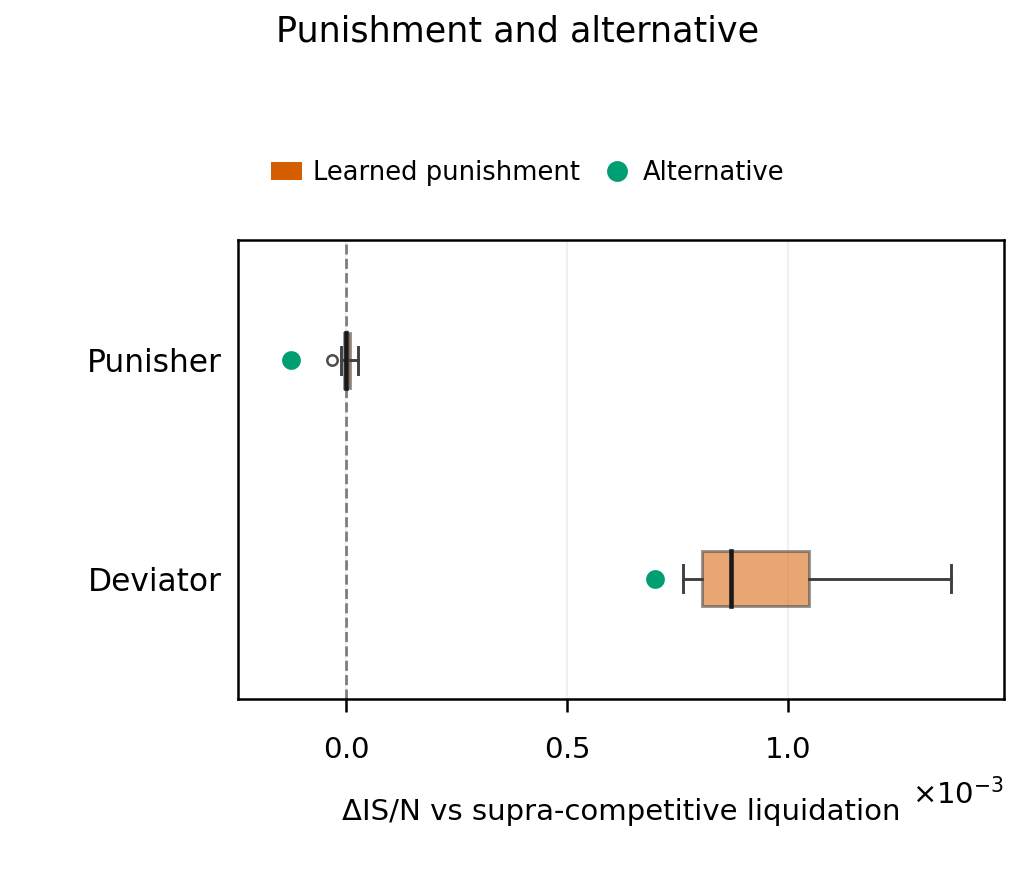}
  \caption{Execution-cost differences given the same forced deviation: learned punishment minus no punishment (orange), and sampled alternative minus no punishment (green).Positive values indicate a loss; negative values indicate a gain.}
  \Description{The learned response has a near-zero cost effect on the punisher and a positive cost effect on the deviator. The alternative schedules lower both players' costs relative to learned punishment, while the deviator's cost remains above the replay control.}
  \label{fig:punishment-costs}
\end{figure}

To quantify the effect of punishment, we impose the same first-step deviation in two cases: in the first, the punisher follows its learned response; in the second, it does not punish and instead follows its learned supra-competitive liquidation path. The deviator resumes its learned policy from timestep 2 in both cases. The orange boxes in Figure~\ref{fig:punishment-costs} report, for each player, the difference in whole-episode execution cost: $\mathrm{IS}/N$ with punishment minus $\mathrm{IS}/N$ without punishment, given the same forced deviation. Since lower execution cost means a higher payoff, a positive difference indicates a loss from punishment and a negative difference indicates a gain.

The orange distributions show that the punisher's average payoff remains materially unchanged compared with not punishing, whereas the deviator incurs a higher cost in every trained run. The punisher thus learns to impose a loss on the deviator without materially reducing its own payoff, preserving on average the payoff from its supra-competitive liquidation path under the same deviation.

This finding contrasts with canonical public-goods experiments, where players pay a direct monetary cost to punish others~\cite{fehr2000cooperation,fehr2002altruistic}. In classical models of collusive price wars, punishment phases also reduce firms' profits relative to continued cooperation~\cite{green1984noncooperative}. Here, the punisher penalises an opponent's deviation from the supra-competitive schedule while managing to preserve the average payoff it would obtain by maintaining its own supra-competitive liquidation path under the same deviation.

The green dots show the mean cost differences across the ten runs when the punisher follows a sampled alternative liquidation trajectory instead of its learned response. We sample a trajectory for the punisher from thousands of nearby feasible schedules, allowing changes to the punisher's trades, excluding its first trade, and enforcing full liquidation at the final timestep. The deviator executes the same forced first trade and is then free to adjust through its learned policy. These sampled trajectories illustrate that the punisher could obtain a higher payoff while imposing a smaller loss on the deviator, yet its learned policy produces stronger punishment rather than choosing the more profitable alternative. 

We next examine whether the deviation and subsequent selling response studied above also occur during training. Across the ten original training runs, we identify episodes in which either player's first trade lies within $0.01$ units of $d_1$, the first action of the learned deviation trajectory. The agents choose these actions during training without intervention. We count each episode only once.

Similarly, we look for additional selling by the other player at timestep 2 that resembles the punitive responses previously identified in our intervention experiments.

\begin{figure}[tb]
  \centering
  \includegraphics[width=\columnwidth]{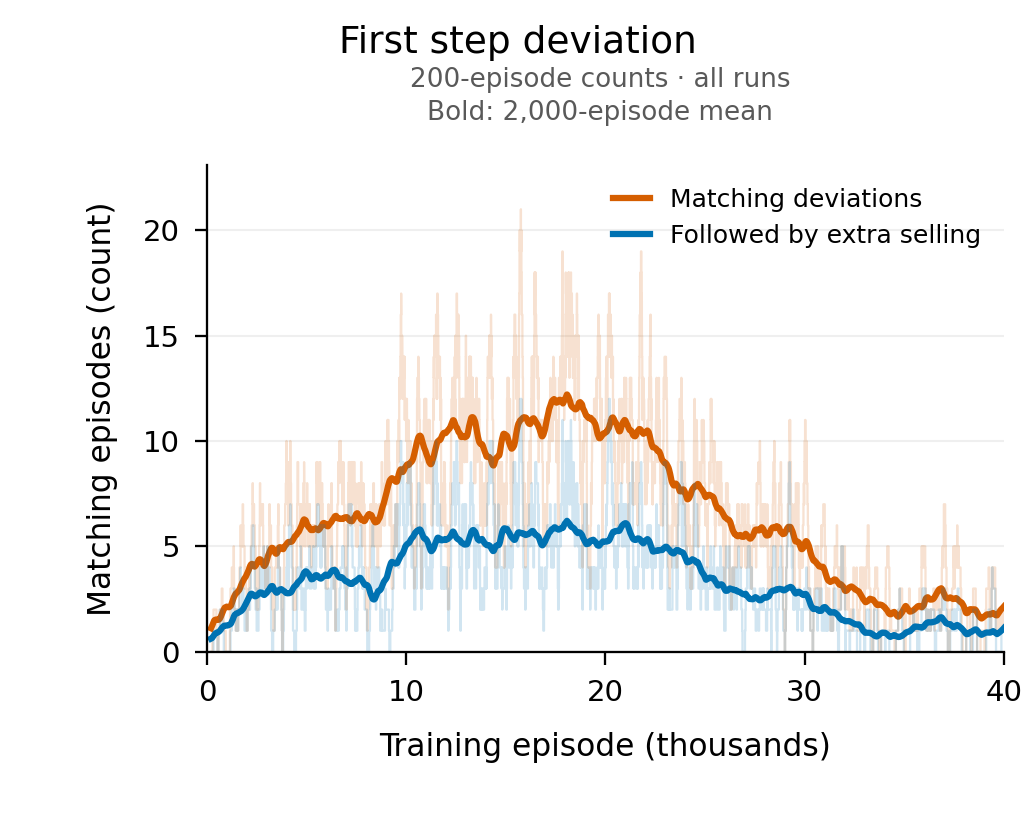}
  \caption{Occurrence of the tested first-step deviation during training. Orange counts matching episodes; blue counts the subset followed by additional selling from the other player. Counts pool a 200-episode trailing window from each of ten runs. Faint curves show rolling counts; bold curves are centred averages over 2,000 successive window endpoints.}
  \Description{Matching deviations and the subset followed by additional selling rise during training, peak near episode eighteen thousand, and then decline.}
  \label{fig:training-deviation-frequency}
\end{figure}

Figure~\ref{fig:training-deviation-frequency} shows that deviation matches initially become more frequent, peak around episode 18,000, and then decline. The subset followed by additional selling exhibits a similar pattern. Together with the intervention results, this evolution is consistent with a learning process in which deviations elicit punitive responses and become less frequent as the supra-competitive outcome stabilises. In this interpretation, successful deterrence reduces the occasions on which punishment is observed.

\subsection{Incentive and Behavioural Checks for Punishment}
\label{sec:punishment-checks}

To give the punitive interpretation an explicit economic basis, we adapt the incentive and behavioural-distance checks of Sivachandran and Paleja~\cite{sivachandran2026curb} to the finite-horizon execution game. Their repeated-game analysis links deterrence to two requirements: punishment must offset the gain from deviation, and the punisher's behaviour must change sufficiently to generate the required loss. We introduce execution-specific counterparts, deriving the behavioural bound directly from our price-impact model. These checks concern the first-step deviation studied above.

\paragraph{The loss from punishment offsets the deviation gain.}
Let $C_B$ denote the deviator's whole-episode $\mathrm{IS}/N$ without a forced deviation, $C_0$ its cost when it deviates and the other player follows its unforced supra-competitive path, and $C_R$ its cost under the same deviation when the other player follows its learned punitive response. The deviator resumes its learned policy after the first trade in both deviation branches. Comparisons use the same initial state. Define
\begin{equation}
  G=C_B-C_0,\qquad H=C_R-C_0.
  \label{eq:deviation-gain-punishment-loss}
\end{equation}
Thus, $G>0$ is the gain available without punishment, and $H>0$ is the loss caused by enabling punishment. The tested deviation is deterred on average when
\begin{equation}
  \mathbb E H\ge\mathbb E G
  \quad\Longleftrightarrow\quad
  \mathbb E C_R\ge\mathbb E C_B.
  \label{eq:execution-incentive-condition}
\end{equation}
Here, expectations are evaluated using averages of paired test cases. Unlike an immediate trading reward, $G$ includes the entire liquidation episode. Because the two deviation branches have identical first trades, their cost difference $H$ arises entirely over timesteps 2--10.

\begin{figure}[tb]
  \centering
  \includegraphics[width=\columnwidth]{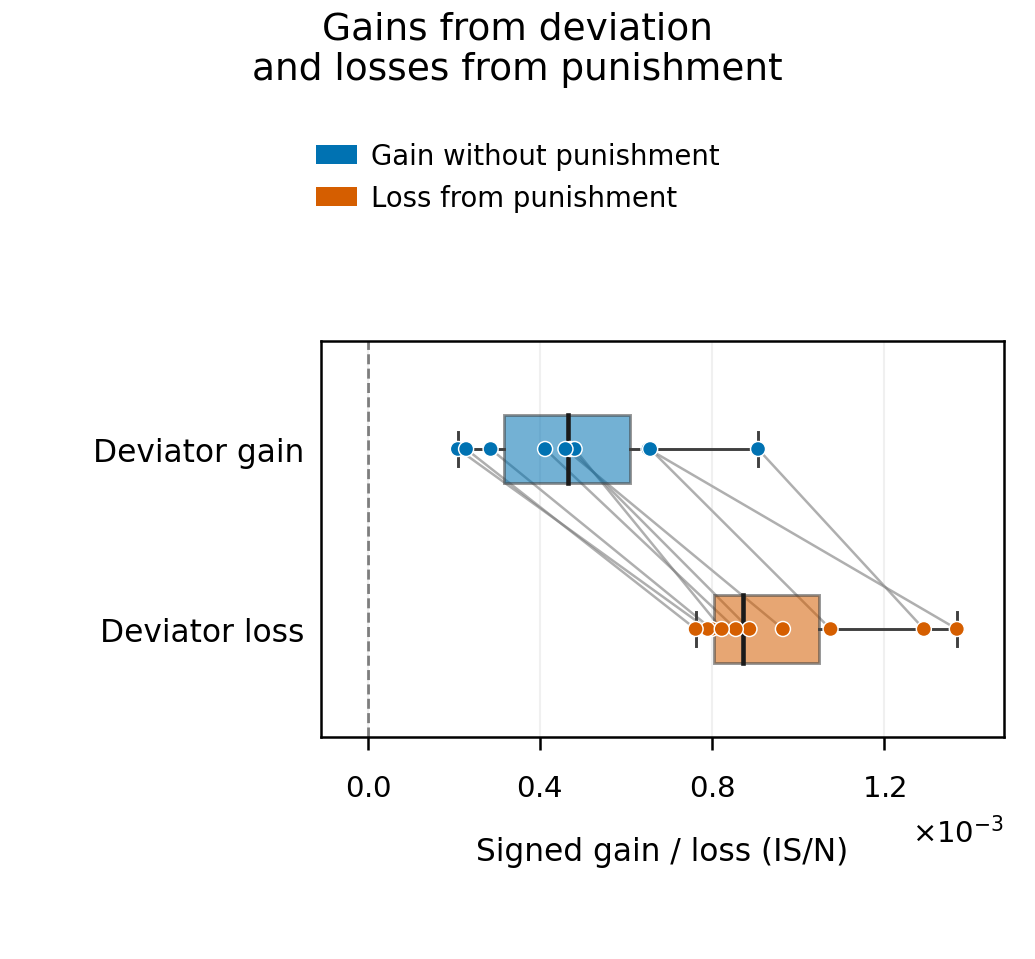}
  \caption{Deviator gain without punishment, $G$ (blue), and loss from punishment, $H$ (orange), in $\mathrm{IS}/N$. Connected points pair values from the same trained run, averaging both player identities and all test episodes. Boxes summarise the ten run means. The punishment loss exceeds the gain in every run.}
  \Description{For every trained run, the orange loss point lies to the right of its connected blue gain point.}
  \label{fig:deviation-gain-punishment-loss}
\end{figure}

Figure~\ref{fig:deviation-gain-punishment-loss} shows that punishment more than offsets the deviation gain in every trained run. Moreover, $G>0$ and $H>G$ hold in all 10,000 paired cases: ten trained pairs, both assignments of the deviator's identity, and 500 test episodes per assignment. The result therefore holds in each direction of deviation, not only after pooling the players.

\paragraph{The change in liquidation timing is sufficiently large.}
Write $x^0,y^0$ for the deviator's and punisher's realised trades without punishment, and $x^R,y^R$ for their trades with punishment, always conditional on the same forced deviation. Let $Q_j=q_0^{(j)}$ be the punisher's initial inventory. The recorded punisher schedules contain only nonnegative sales and both liquidate $Q_j$ fully. We can therefore measure the fraction of inventory redistributed across trading times by
\begin{equation}
  \mathrm{TV}_{\mathrm{time}}
  =\frac{1}{2Q_j}\sum_{s=1}^{N}|y_s^R-y_s^0|.
  \label{eq:liquidation-timing-tv}
\end{equation}
This is total variation between distributions of liquidation volume over timesteps, rather than between the conditional action distributions in the motivating repeated-game analysis.

The deviator may also adjust its later trades. Let $\mathcal C(x,y)$ denote its realised $\mathrm{IS}/N$ for recorded schedules $x,y$, and define
\begin{equation}
  A=\mathcal C(x^R,y^R)-\mathcal C(x^0,y^R).
  \label{eq:deviator-adaptation}
\end{equation}
This term measures the cost effect of the deviator's adjustment while keeping the punisher's trades fixed at $y^R$. The counterfactual cost $\mathcal C(x^0,y^R)$ combines the deviator's recorded trades without punishment with the punisher's recorded trades under punishment, using the same realisation of price noise. Both trade sequences are held fixed in this calculation, they need not arise jointly when the agents follow their learned policies.

A changed punisher sale at timestep $s$ affects the deviator's current execution price through temporary impact and its later prices through permanent impact. Define the corresponding weights and their range by
\begin{align}
  w_s&=\frac{a}{\tau}x_s^0+\kappa\sum_{t>s}x_t^0,\nonumber\\
  \Omega&=\max_{2\le s\le N}w_s-\min_{2\le s\le N}w_s.
  \label{eq:execution-exposure-range}
\end{align}
The range includes final liquidation. To derive the bound, set $z_s=y_s^R-y_s^0$. In deterministic evaluation, simultaneous decisions and identical initial observations give $z_1=0$, while full liquidation gives $\sum_s z_s=0$. Under the same realisation of price noise, Equations~\eqref{eq:execution-price}--\eqref{eq:midprice-dynamics} yield
\begin{equation}
  H-A=\frac{1}{N}\sum_{s=2}^{N}w_s z_s.
  \label{eq:direct-punishment-cost}
\end{equation}
Subtracting the midpoint of the largest and smallest weights leaves this sum unchanged, since the trade differences sum to zero. Each centred weight then has absolute value at most $\Omega/2$. The triangle inequality gives
\begin{equation}
  |H-A|\le\frac{\Omega}{2N}\sum_{s=1}^{N}|z_s|
  =\frac{Q_j\Omega}{N}\mathrm{TV}_{\mathrm{time}}.
  \label{eq:execution-tv-cost-bound}
\end{equation}
Averaging and imposing~\eqref{eq:execution-incentive-condition} therefore requires
\begin{equation}
  \frac{Q_j}{N}\mathbb E[\Omega\,\mathrm{TV}_{\mathrm{time}}]
  \ge[\mathbb E G-\mathbb E A]_+,
  \label{eq:execution-necessary-magnitude}
\end{equation}
where $[u]_+=\max(u,0)$. This condition states that the change in liquidation timing must be large enough to offset the deviation gain after accounting for the deviator's subsequent adjustment.

To display both sides in TV units, divide by $(Q_j/N)\mathbb E\Omega$, which is positive in our data. The equivalent comparison is
\begin{equation}
  \underbrace{\frac{\mathbb E[\Omega\,\mathrm{TV}_{\mathrm{time}}]}{\mathbb E\Omega}}_{\mathrm{TV}_{\Omega}:\ \text{measured}}
  \ge
  \underbrace{\frac{N[\mathbb E G-\mathbb E A]_+}{Q_j\,\mathbb E\Omega}}_{b_{\Omega}:\ \text{required}}.
  \label{eq:weighted-tv-condition}
\end{equation}
The measured value weights each case's timing change by its price-impact exposure $\Omega$; it is not the TV of an averaged trajectory. Products are averaged before division.

\begin{figure}[tb]
  \centering
  \includegraphics[width=\columnwidth]{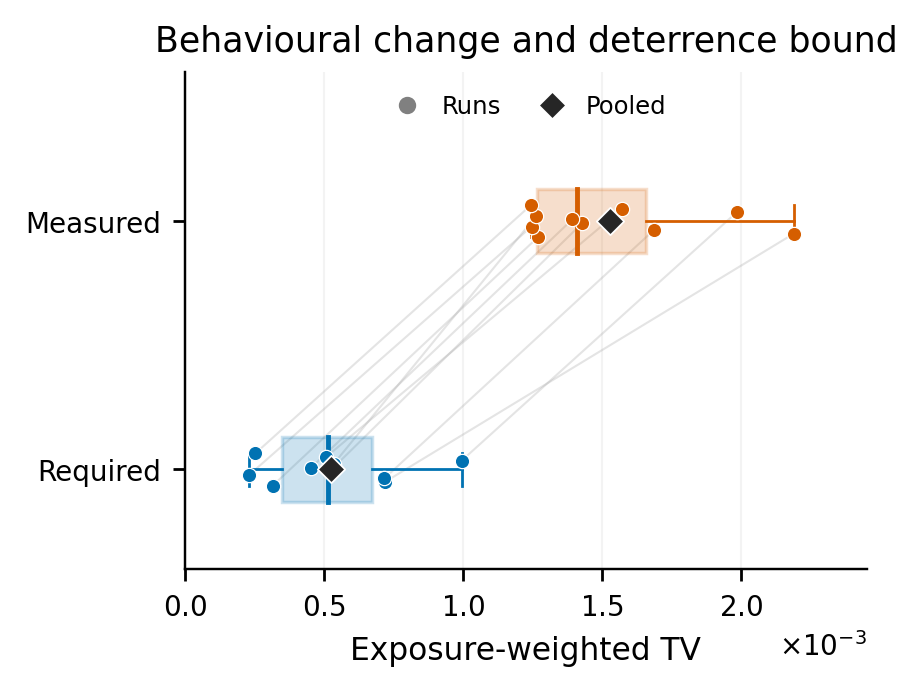}
  \caption{Measured exposure-weighted TV and its required boundary from~\eqref{eq:weighted-tv-condition}. Connected points compare the two quantities within each trained run; boxes summarise their distributions and black diamonds show pooled values. The measured change exceeds the boundary in every run.}
  \Description{The measured timing change lies above its required boundary in all ten trained runs and in the pooled comparison.}
  \label{fig:tv-deterrence-bound}
\end{figure}

Figure~\ref{fig:tv-deterrence-bound} shows that measured TV exceeds the required boundary in every run. The corresponding case-level magnitude condition also holds in all 10,000 paired evaluations. Thus, both checks pass for the tested one-step deviation: the learned punishment eliminates its profitability, and the change in liquidation timing exceeds the necessary magnitude. 

\section{Conclusions}
\label{sec:conclusions}

In this paper, we build on the optimal-execution studies of Lillo and Macr\`i~\cite{lillo2026deviations} and Koulouris and Campajola~\cite{koulouris2026memory} by identifying punitive responses that provide a behavioural basis for interpreting supra-competitive outcomes as collusive. We examine a two-player, finite-horizon Almgren--Chriss liquidation game using independent PPO agents with access to within-episode price and action histories. Across all ten trained pairs, both agents achieve execution costs below the Nash benchmark, without explicit communication or a shared reward. Training against a fixed version of the learned schedule identifies a profitable unilateral deviation, providing an economically meaningful probe of the original policies.

When the deviation is imposed at the first timestep, the other agent accelerates liquidation and increases the deviator's execution cost. The loss imposed exceeds the gain available without punishment, while the punisher's average payoff remains materially unchanged relative to maintaining its supra-competitive path under the same deviation. Nearby sampled liquidation plans offer the punisher higher payoffs while imposing a smaller loss on the deviator. The learned interaction therefore exhibits punishment that deters the tested deviation without a material sacrifice of the punisher's own payoff.

The training trajectories reveal a complementary temporal pattern (Figure~\ref{fig:training-deviation-frequency}): matches to the tested deviation and subsequent additional selling first become more frequent, then decline towards convergence. The final-policy interventions show that this decline coexists with an effective punitive response when the deviation is imposed again. Together, these observations are consistent with deterrence becoming established: the tested deviation pattern becomes less frequent, while a renewed deviation still triggers a response that removes its profitability.

We give this interpretation an explicit economic basis through the incentive comparison and an execution-specific total-variation bound linking changes in liquidation timing to their price-impact consequences. Both checks hold for the tested first-step deviation across all runs and both player identities. Together, these findings support a collusive interpretation of the learned interaction beyond the observation of supra-competitive costs alone. Future work should examine whether the mechanism persists under stronger price noise, larger agent populations and a broader range of deviations.

\bibliography{references}
\end{document}